\documentclass[11pt,twoside]{article} 
\usepackage{asp2004}
\usepackage{epsf}
\usepackage{psfig}
\usepackage{lscape} % for landscape figures/tables

\begin{document} 
\title{Subdwarf B Binaries from the Edinburgh--Cape Survey}
\author{L. Morales-Rueda,$^1$ P.F.L. Maxted,$^2$ T.R. Marsh,$^3$
D. Kilkenny,$^4$ and D. O'Donoghue$^4$} \affil{$^1$ Department of
Astrophysics, University of Nijmegen, P.O. Box 9010, 6500 GL,
Nijmegen, NL\\ $^2$ Astrophysics group, Keele University, Keele,
Staffordshire ST5 5BG, UK\\ $^3$ Department of Physics, Warwick
University, Coventry CV4 7AL, UK\\ $^4$ SAAO, PO Box 9, Observatory,
7935, ZA}

\begin{abstract} 
We present the first results of a campaign to obtain orbital solutions
of subdwarf B (sdB) stars from the Edinburgh-Cape survey. We have
obtained blue spectra of 35 sdBs, 20 of which have been observed in
more than two epochs. 15 out of the 35 are certain binaries with a few
other objects showing radial velocity variations with small amplitude,
possibly long period sdB binaries. We have secured the orbital
parameters for 2 of the 15 systems and narrowed down the orbits of
another one to a small range of periods. These preliminary results
only use data taken up to December 2003.
\end{abstract}

\section{Introduction}
In March 2002 we started a campaign to obtain orbital solutions of 100
subdwarf B (sdB) binaries from the Edinburgh-Cape (EC) survey (Stobie
et al. 1997). This work was intended as the Southern Hemisphere
counterpart of that undertaken by Maxted et al. (2001) and
Morales-Rueda et al. (2003). In their work Morales-Rueda et al. (2003)
select their targets from two main sources, the Palomar-Green (PG) and
the Kitt Peak-Downes (KPD) catalogues. The PG catalogue is biased
against targets that show a Ca II H-line (as these were taken out of
the catalogue), and thus, against sdB binaries with main sequence
companions. On the other hand the EC survey does not show these
biases, as the object selection is based on colours only, and is ideal
to obtain an ``unbiased'' sample of sdB binaries that can then be
compared with binary evolution theories (Han et al. 2003). We should
mention that some bias might still remain as in the cases where the
main sequence star was somewhat brighter than the sdB star
(i.e. combined colour redder than $U-B$ = $-$0.4 ), the binary would
have been classified probably as a late A star and would have been
taken out of the catalogue.

\section{Observations, Data Reduction and Period Determination}

The spectra were obtained with the grating spectrograph plus the SITe
CCD at the 1.9\,m Radcliffe telescope at the SAAO. Grating 4, with
1200 grooves per mm was used to obtain spectra covering H$\beta$ and
H$\gamma$ with a dispersion of 0.5\,\AA/pix and a resolution of less
than 1\,\AA\ at 4600\,\AA. The spectra were reduced according to
standard procedures. The Balmer lines were fitted with a model line
profile consisting of three Gaussians to measure their radial
velocities. These were then fitted with functions consisting of a
sinusoidal plus a constant with 4 free parameters, i.e. the constant,
the semiamplitude of the sinusoidal, the period and the zero point. A
value of the $\chi^2$ of each fit was obtained for each combination of
parameters and a periodogram of the form shown in the right hand
panels of Figs. 1, 2 and 3, obtained for each system. We considered
that the orbital period of the binary was determined when the
difference in $\chi^2$ between the first and second alias was larger
than 20. We calculated the probability that the true orbital period
lies further than 1 and 10 per cent from our chosen period and present
these values in Table 1 together with the orbital solution for each
system. For details on the data reduction, period determination and
probability calculations see Morales-Rueda et al. (2003).

\section{Results: Orbital Parameters for Three New sdB Binaries}

We find the orbital solutions for two systems EC00404$-$4429 and
EC02200$-$2338. These are presented in Table 1 and in Figs. 1 and 2
respectively. For another system, EC12327$-$1338, we find that the
orbital period lies around 0.36 d but the number of observations is
not enough to obtain an accurate orbital period.

\begin{table}[!ht]
\caption{Orbital solutions for three sdB binaries from the EC
  survey. In the case of EC12327$-$1338 a range of parameters is given
  as there are 13 aliases in a small range of periods. $\Delta\chi^2$
  is the difference in $\chi^{2}$ between the 1st and 2nd aliases. n
  is the number of radial velocities used for the orbit
  determination. The rows labelled as {\em 1 and 10 per cent} give the
  probabilities that the true period lies further than 1 and 10 per
  cent from our favoured value. Numbers given are the $\log_{10}$ of
  the probabilities. The last row gives the systematic uncertainty
  that has been added in quadrature to the raw error to give a
  $\chi^{2}$ that lies above the 2.5 per cent probability in the
  $\chi^{2}$ distribution. This systematic uncertainty accounts for
  unaccounted sources of error like slit-filling, intrinsic
  variability of the star etc. Numbers marked with a $^*$ have been
  calculated for the lowest $\chi^{2}$ alias, P = 0.36397(4) d.}
\smallskip
\begin{center}
{\small
\begin{tabular}{lccc}
\tableline
\noalign{\smallskip}
 & EC00404$-$4429 & EC02200$-$2338 & EC12327$-$1338\\
\noalign{\smallskip}
\tableline
\noalign{\smallskip}
Period [d] & 0.12834(4) & 0.8022(7) & 0.3628 - 0.3674\\%0.3651(23)\\
HJD$_0$ [d] & 2452895.4418(4) & 2452896.029(4) & \\
$\gamma$ [km s$^{-1}$] & 32.97 $\pm$ 2.94 & 20.74 $\pm$ 2.29 & $-$7.43
- $-$4.06\\
K [km s$^{-1}$] & 152.8 $\pm$ 3.4 & 96.35 $\pm$ 1.43 & 119.8 - 124.3\\
M$_2$min [M$_{\odot}$] & 0.32 & 0.39 & 0.38$^*$\\
$\chi^{2}_{reduced}$ & 1.3 & 0.42 & 1.6-3.6\\
2nd best alias [d] & 0.11350(3) & 0.3038(1)\\
$\Delta\chi^2$  & 73 & 61\\
n & 9 & 10 & 13\\
1 per cent & -15.18 & -11.35 & -3.47$^*$\\
10 per cent & -15.18 & -11.35 & -10.81$^*$\\
Systematic error & 2 & 2 & 2\\
\noalign{\smallskip}
\tableline
\end{tabular}
}
\end{center}
\end{table}

\begin{figure}[!ht]
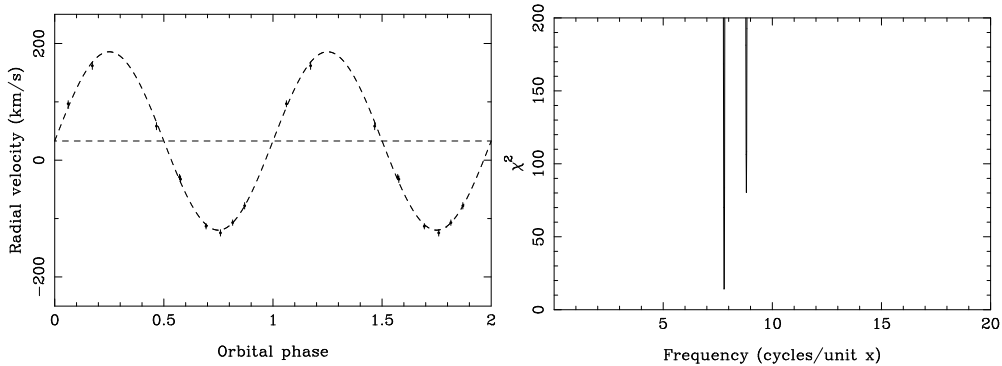

\plotfiddle{ec00404-4429pfold.ps}{1.5cm}{-90}{28}{28}{-195}{80}
\plotfiddle{ec00404-4429pgram.ps}{0cm}{-90}{28}{28}{-5}{103}
\vspace{18mm}
\caption{{\bf EC00404$-$4429:} Left panel: radial velocity curve
solution fitted to the data measurements. Right panel: periodogram
showing the favoured period and the closest aliases.}
\end{figure}

\begin{figure}[!ht]
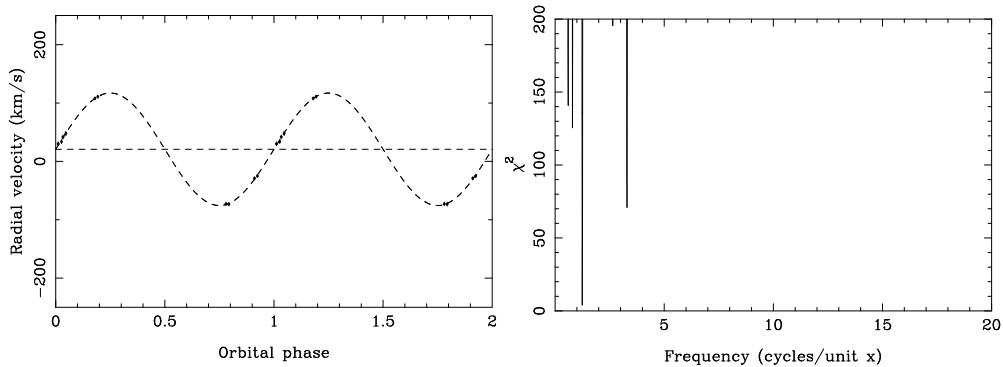

\plotfiddle{ec02200-2338pfold.ps}{1.5cm}{-90}{28}{28}{-195}{80}
\plotfiddle{ec02200-2338pgram.ps}{0cm}{-90}{28}{28}{-5}{103}
\vspace{18mm}
\caption{{\bf EC02200$-$2338:} Same as in Fig.1}
\end{figure}

\begin{figure}[!ht]
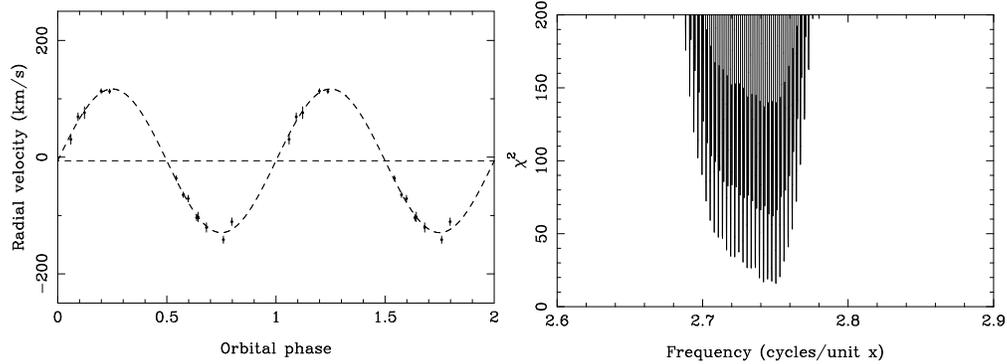

\plotfiddle{ec12327-1338pfold.ps}{1.5cm}{-90}{28}{28}{-195}{80}
\plotfiddle{ec12327-1338pgramB.ps}{0cm}{-90}{28}{28}{-5}{103}
\vspace{18mm}
\caption{{\bf EC12327$-$1338:} Same as in Fig.1. The data have been
  folded using the orbit solution with smaller $\chi^{2}$. The
  periodogram shows many aliases comprised in the range given in
  Table 1.}
\end{figure}

\section{Discussion}

Their orbital periods place EC00404$-$4429, EC02200$-$2338 and
 EC12327$-$1338 in the group of sdB binaries formed via the common
 envelope (CE) ejection channel (see fig. 10 of Han et al. 2003). The
 minimum mass of the companion, calculated assuming the canonical mass
 of 0.5\,M$_{\odot}$ for the sdB, indicates that the companions are
 probably white dwarfs, thus these binaries have probably formed
 through the second CE path (Han et al. 2003).

We have added the orbital periods of these three new systems to the
orbital period distribution of sdBs known, Fig. 4. The most
interesting features of this distribution are the excess of binaries
at orbital periods around 1 d and that the very long period systems
(tens and hundreds of days) are starting to appear in the sample. The
comparison of the observed distribution with theoretical predictions
(Han et al. 2003) is invaluable to determine, among other important
evolutionary unknowns, plausible values for the CE ejection
efficiency.

\begin{figure}[!ht]
\plotfiddle{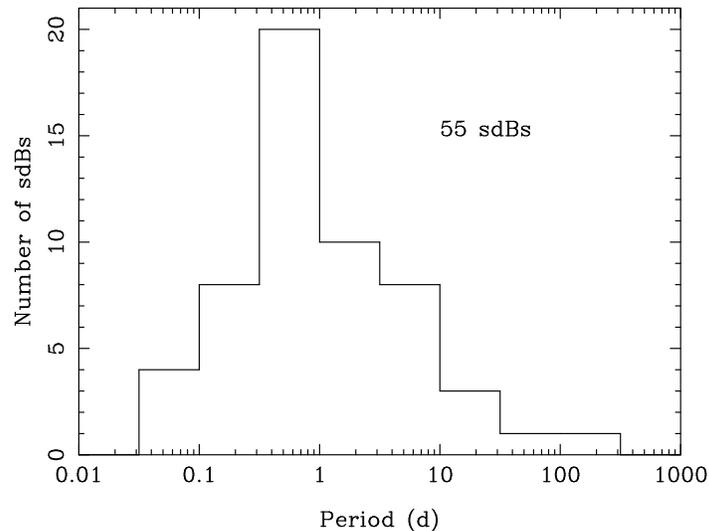}{2.5cm}{-90}{40}{40}{-150}{120}
\vspace{35mm}
\caption{Orbital period distribution for the sdB binaries known. The
  figure is an update of Morales-Rueda et al. (2003) and Morales-Rueda
  et al. (2004).}
\end{figure}

\acknowledgements{LMR was supported by a PPARC postdoctoral grant
during this research. This paper uses observations made at the South
African Astronomical Observatory (SAAO).}

\end{document}